\documentclass[%
 aip,
 jcp,
 reprint%
]{revtex4-1}

\usepackage{amsmath}
\usepackage{amssymb}
\usepackage{graphicx}% Include figure files
\usepackage{dcolumn}% Align table columns on decimal point
\usepackage{bm}% bold math
\usepackage[utf8]{inputenc}
\usepackage[T1]{fontenc}
\usepackage{mathptmx}
\usepackage{etoolbox}
\usepackage{csquotes}

\usepackage{siunitx}
\usepackage{braket}
\usepackage[version=4]{mhchem}

\usepackage{float}

\makeatletter
\def\@email#1#2{%
 \endgroup
 \patchcmd{\titleblock@produce}
  {\frontmatter@RRAPformat}
  {\frontmatter@RRAPformat{\produce@RRAP{*#1\href{mailto:#2}{#2}}}\frontmatter@RRAPformat}
  {}{}
}%
\makeatother
\begin{document}

\preprint{AIP/123-QED}

\title{Simulating Ultrafast Transient Absorption Spectra from First Principles using a Time-Dependent Configuration Interaction Probe}
% Force line breaks with \\
\author{Arshad Mehmood}
\affiliation{Department of Chemistry, 
Stony Brook University, Stony Brook, New York 11794, USA}
\affiliation{Institute for Advanced Computational Science, 
Stony Brook University, Stony Brook, New York 11794, USA}

\author{Myles C. Silfies}
\altaffiliation[Current address: ]{Space Dynamics Laboratory, North Logan, Utah 84341, USA}
\affiliation{Department of Physics, 
Stony Brook University, Stony Brook, New York 11794, USA}

\author{Andrew S. Durden}
\altaffiliation[Current address: ]{Department of Chemistry, Wayne State University, Detroit, Michigan 48202, USA}
\affiliation{Department of Chemistry, 
Stony Brook University, Stony Brook, New York 11794, USA}
\affiliation{Institute for Advanced Computational Science, 
Stony Brook University, Stony Brook, New York 11794, USA}

\author{Thomas K. Allison}
\affiliation{Department of Chemistry, 
Stony Brook University, Stony Brook, New York 11794, USA}
\affiliation{Department of Physics, 
Stony Brook University, Stony Brook, New York 11794, USA}

\author{Benjamin G. Levine}
\affiliation{Department of Chemistry, 
Stony Brook University, Stony Brook, New York 11794, USA}
\affiliation{Institute for Advanced Computational Science, 
Stony Brook University, Stony Brook, New York 11794, USA}
\email{ben.levine@stonybrook.edu}
%\phone{+1 631-632-2381}
%\fax{+1 631-632-7960}

%\author{A. Author}
%\author{B. Author}%
% \email{Second.Author@institution.edu.}
%\affiliation{ 
%Authors' institution and/or address%\\This line break forced with \textbackslash\textbackslash
%}%

% \homepage{http://www.Second.institution.edu/~Charlie.Author.}
%\affiliation{%
%Second institution and/or address%\\This line break forced% with \\
%}%

\date{\today}% It is always \today, today,

\begin{abstract}
Our results are compared to gas-phase TAS data recorded from both jet-cooled ($T\sim 40$ K) and hot ($\sim 403$ K) molecules via cavity-enhanced transient absorption spectroscopy (CE-TAS).
Decomposition of the computed spectrum allows us to assign a rise in the SE signal to excited-state proton transfer and the ultimate decay of the signal to relaxation through a twisted conical intersection.
The total cost of computing the observable signal ($\sim$1700 graphics processing unit hours for $\sim$4 ns of electron dynamics) was markedly less than that of the {\em ab initio} multiple spawning calculations used to compute the underlying nonadiabatic dynamics.

\end{abstract}

\maketitle

\section{Introduction}
The advent of the ultrafast laser pulse transformed the way we learn about the dynamics of molecules.\cite{Zewail2000}  Prior to this development, the nature of short-lived chemical species could only be inferred from long-time outcomes.  But with ultrafast laser pulses, it was possible to take a spectroscopic picture of short-lived species, such as transition states.\cite{Polanyi1995,Bowman1989}  Yet ultrafast spectra remain very difficult to interpret.  What one wishes to learn from an ultrafast experiment is the time-dependent molecular wave function, $\ket{\Psi(t)}$, but what one actually measures is a lossy projection, whose information content is determined by the nature of the probe pulse and measured signal.  For a given method, some features of $\ket{\Psi(t)}$ will be clearly discerned, while others will be effectively invisible.

As such, ultrafast theory has become an essential partner to experiment.  In the late 1990s, pioneering work on the solvated electron,\cite{Schwartz1994a,Schwartz1994b,Schwartz1994c,Schwartz1994d} photodissociation,\cite{Batista1997} and photoisomerization\cite{Vreven1997,BenNun1998} demonstrated the utility of ab initio molecular dynamics (AIMD) for the simulation of ultrafast dynamics. 
The AIMD approach is to run molecular dynamics simulations in which the PESs are computed via on-the-fly electronic structure calculations at each time step. 
Using AIMD, PESs of relatively large molecules may be explored in their full dimensionality, with minimal prior knowledge of the PES and/or reaction coordinate.

In the end, AIMD provides the researcher with an approximate time-dependent wave function, $\ket{\Psi_{\text{approx}}(t)}$, that can be used to assign the features of the experimental spectrum.  But the fact that $\ket{\Psi_{\text{approx}}(t)}$ is approximate makes assignment challenging.  While experiment provides incomplete information about the exact physical system of interest, simulation provides relatively complete information about an inexact system.  A common procedure for making such assignments is to look for processes in the simulation data whose timescales correspond roughly to lifetimes observed in the time-dependent experimental signal.  But as will be exemplified in this work, assigning spectra using simulated lifetimes can be fraught.  All simulations are based on approximate Hamiltonians, which yield approximate energies. 
Particularly for statistical dynamics,\cite{Steinfeld_Book1999} where rates are related to the exponential of a barrier height, a small error in the relevant energy difference may be amplified significantly.  
For example, an error of 1 kcal/mol in a barrier height (commonly considered to be "chemical accuracy" in the theory community) may translate to an error of a factor of five in a predicted lifetime at room temperature, and more at the cold temperatures at which many gas phase experiments are carried out.  Therefore, lifetimes are arguably among the most error-prone quantities that can be derived from a dynamical simulation, and therefore a poor choice to guide assignments.

To make more definitive assignments, one may compute a less error-prone observable for comparison, such as the time-dependent spectrum itself.  
Direct comparison of the observable measured by the probe pulse removes much of the ambiguity inherent to assigning spectra, with the assignment now based on unique spectral fingerprints for the contributing states and/or geometries.
In previous AIMD work, many have done just this.  For example, time-resolved photoelectron spectroscopy (\mbox{TRPES}) was one of the earliest ultrafast spectra to be directly computed from a swarm of AIMD simulations.\cite{Hudock2007}  This was a natural early target, and has become a mainstay of computational ultrafast spectroscopy,\cite{Mitric2011,Dsouza2018,Chakraborty2022} because the loose selection rules of the \mbox{TRPES} probe allow the observation of a wide range of processes, and the absence of solvent makes computation relatively straight forward.  

Ultrafast electron diffraction (UED), which can be computed directly from interatomic distances without additional electronic structure calculations, provides a tight and convenient connection between experiment and theory.\cite{STEFANOU2017300,Wolf2019,parrish2019,champenois2021aims,liu2023rehybridization}  In fact, a community challenge is testing the extent to which theory can predict the signal of a UED experiment, {\em a priori}.\cite{Miao2024,Santa2024,Suchan2024,Janos2024,Eng2024,Mukherjee2024}\footnote{These references will be updated at the revision stage to include not-yet-published responses to the challenge.}  Though the experimental result remains unannounced, it is already clear that the lifetimes predicted by the simulations range widely.  
In the fastest simulations, decay to the electronic ground state is observed in $\sim$100 fs, but in the slowest the population remains entirely in the electronic excited state beyond 1 ps.  
Despite this variability in lifetime, the diffraction signal associated with ring opening is relatively consistent from prediction to prediction, which will aid in assigning experimental features. 
This example supports the assertion that direct comparison of simulated observables is less error prone than the comparison of lifetimes alone.

Observables more directly sensitive to electronic structure remain quite difficult both to interpret and to simulate.  Despite being the most commonly used ultrafast spectroscopy method, transient absorption spectroscopy (TAS) is in this category.  
In TAS, a pump pulse electronically excites the system of interest, initiating wavepacket motion on the electronic excited state.  
Then, after the wavepacket evolves for a fixed delay time, a second pulse probes the current state of the wavepacket, and the absorbance is measured.  
Transient absorption spectrometers operating in the UV and visible can now routinely achieve $\sim 10$ fs resolution \cite{Cerullo2020} and can also simultaneously achieve high spectral resolution if this is desired or appropriate, since the time resolution of the experiment and the spectral resolution are not conjugate variables.\cite{Pollard_JCP1990}
While TAS is usually applied to condensed-phase systems with high optical density, recent work by Allison and co-workers has used frequency combs and optical resonators to perform cavity-enhanced transient absorption spectroscopy (CE-TAS) with sensitivity several orders of magnitude lower than conventional TAS methods.\cite{Silfies_PCCP2021,Reber_Optica2016,Allison_JPhysB2017}
CE-TAS can be applied to dilute gas-phase molecules in molecular beams, and these gas-phase studies, free of solvent interactions, allow very meaningful comparison to first principles theory. \cite{Silfies2023}  

Despite the label of \enquote{transient absorption,} there are several contributions to the TAS signal.  
In addition to excited-state absorption (ESA), which registers as a positive signal, the total signal also includes stimulated emission (SE) and ground-state bleach (GSB) signals, which both register as negative. 
\textit{A priori} assignment of TAS signals is thus not straightforward, in part because the ESA, SE, and GSB signals may be overlapping and quite broad.  
Just as for \mbox{TRPES} and UED, direct simulation of the TAS probe signal from dynamical simulations offers a path forward.  Direct simulation of TAS signals based on fitted PESs and model Hamiltonians has provided important insights into both the general features observed in experimental TAS spectra \cite{Scholes2016,Domcke2019} and the methodological features required for accurate simulation.\cite{Mukamel1994,Mukamel1995,Cina1999,Stock2000,Geva2009,Vanicek2013,Domcke2013,Subotnik2014}  
Recently, AIMD is also becoming a popular approach to simulating TAS. Most simulations to date have been based on real-time or linear-response time-dependent density functional theory (TDDFT) or density functional tight binding (DFTB), which are widely held to offer a favorable balance of computational cost and accuracy.\cite{Rubio2013,Govind2015,Parkhill2016,Sanchez2018,Sanchez2020,Rubio2020,Lan2021,Garavelli2022} 
Recently there have been several studies in which the probe signal is computed using correlated wave function methods,\cite{Domcke2021,Kubas2021,Silfies2023}, including truly heroic efforts that use highly accurate complete active space second-order perturbation theory (CASPT2) for this purpose.\cite{Cerullo2021,Xu2022}

From an electronic structure perspective, accurately simulating the probe signal remains a significant challenge.  For one, ESA may access many high-energy excited states, some of which may be doubly excited with respect to the ground state.  This renders many standard single-reference methods for computing electronic absorption spectra less than ideal.\cite{Roldao2022}  Yet efficiency is crucial, given that filling in the two-dimensional (probe wavelength vs time) TAS signal requires a very large number of individual electronic structure calculations.  One may need to contend with a high density of electronic states of different electronic character (\textit{e.g.} local, charge transfer, Rydberg), which increases computational cost and may trigger convergence difficulties that are frustrating to solve {\em en masse}.  The human effort required to solve thousands of individual convergence failures could render a study unfeasible, thus a robust approach to simulating the probe pulse is essential.  Finally, the presence of solvent interactions in conventional TAS experiments requires the inclusion of, at the minimum, implicit solvent models.\cite{Cramer1999,Herbert2021,Ringe2022}  From this perspective, the introduction of CE-TAS allows a closer connection between experiment and theory than previously possible, by enabling the accurate calculation of the experimental signal without including solvent effects.   

To address this complexity, in this paper we present a novel method for computing transient absorption spectra from AIMD simulation results.
In order to efficiently and robustly model ESA and SE, the spectrum is computed via time-dependent complete active space configuration interaction (TD-CASCI), which offers several advantages over traditional time-independent electronic structure calculations in this context. 
First is robustness.
Like other real-time electronic structure methods,\cite{Li2020} a TD-CASCI simulation is an initial value problem. 
Unlike the time-independent eigenvalue problem, real-time propagation of the wave function never fails to converge.  Even though the simulation of linear spectra from time-independent simulations is straightforward, real-time approaches are increasingly popular for this purpose given their ability to stably compute a large number of spectral peaks without storing or converging a large number of eigenvectors.\cite{Yabana1999,Marquis2005,Lopata2013,Ranelka2015,Goings2016,nascimento_2016,nascimento_2017,Woicik2020}  Additionally, the CASCI ansatz is capable of describing the double and higher-order excitations that may be accessed upon ESA.  Finally, graphics processing unit (GPU) acceleration enables tens of thousands of spectra to be computed using relatively modest computing resources.  

\begin{figure}[t]
	\includegraphics[width=3.25in]{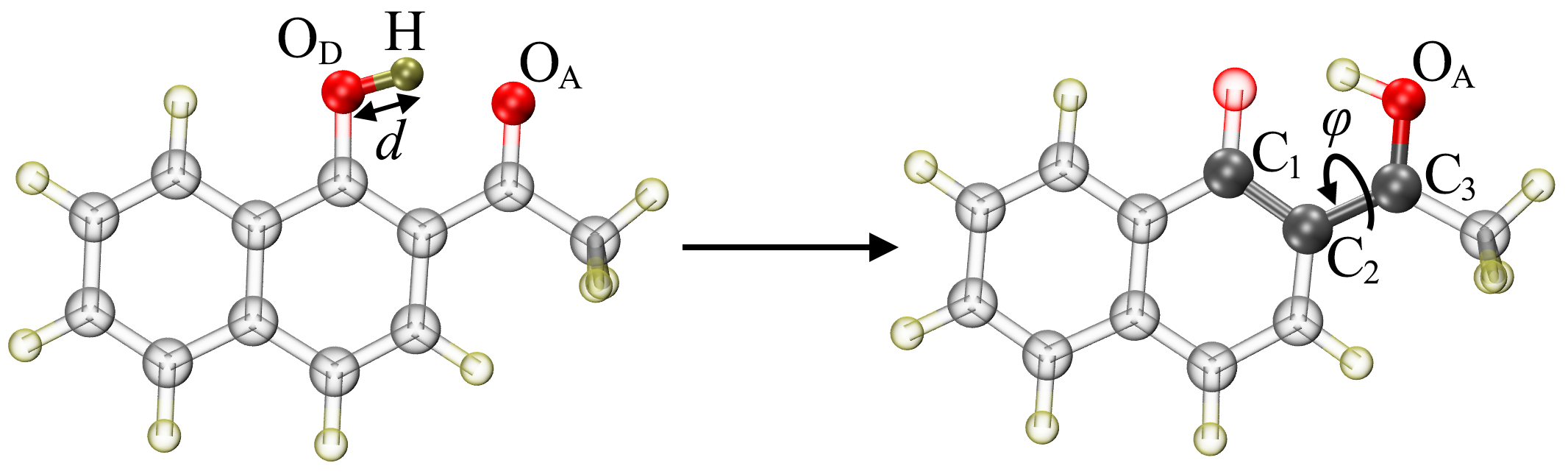}
	\caption{Structure of HAN in the enol (left) and keto (right) tautomers. The ESIPT involves the transfer of H  from donor O\textsubscript{D} to acceptor O\textsubscript{A} oxygen atoms. The dihedral angle  formed by labeled carbon and O\textsubscript{A} atoms controls the lifetime of keto form in the \textit{S}\textsubscript{1} state.}
	\label{fig:enolketo}
\end{figure}

Below we describe this new approach in detail and demonstrate its utility by application to 1$^\prime$-hydroxy-2$^\prime$-acetonaphthone (HAN; Fig. \ref{fig:enolketo}), a prototypical excited-state intramolecular proton transfer (ESIPT) system.    
This paper provides a more detailed presentation of the method than our previous experimentally-focused paper on salicylideneaniline.\cite{Silfies2023} As in that work, our calculated spectra are directly compared to CE-TAS measurements on the gas-phase molecule under a range of initial conditions.
Directly simulating the ultrafast spectroscopic observable enables decomposition of the spectrum into components corresponding to distinct physical processes.
    
In section \ref{sec:theory}, we will describe our TD-CASCI-based approach to computing TAS spectra.  In section \ref{sec:expt}, we present the CE-TAS experiment that serves both the experiment we aim to interpret in this work and as a benchmark for our simulation method.  In section \ref{sec:results}, we compare the experimental and simulated spectra, and decompose the simulated spectrum in order to gain insights in the dynamics of HAN.  In section \ref{sec:conclusions}, we draw conclusions and discuss future prospects.

\section{Theoretical Methods}
\label{sec:theory}

\begin{figure}[t]
	\includegraphics[width=3.25in]{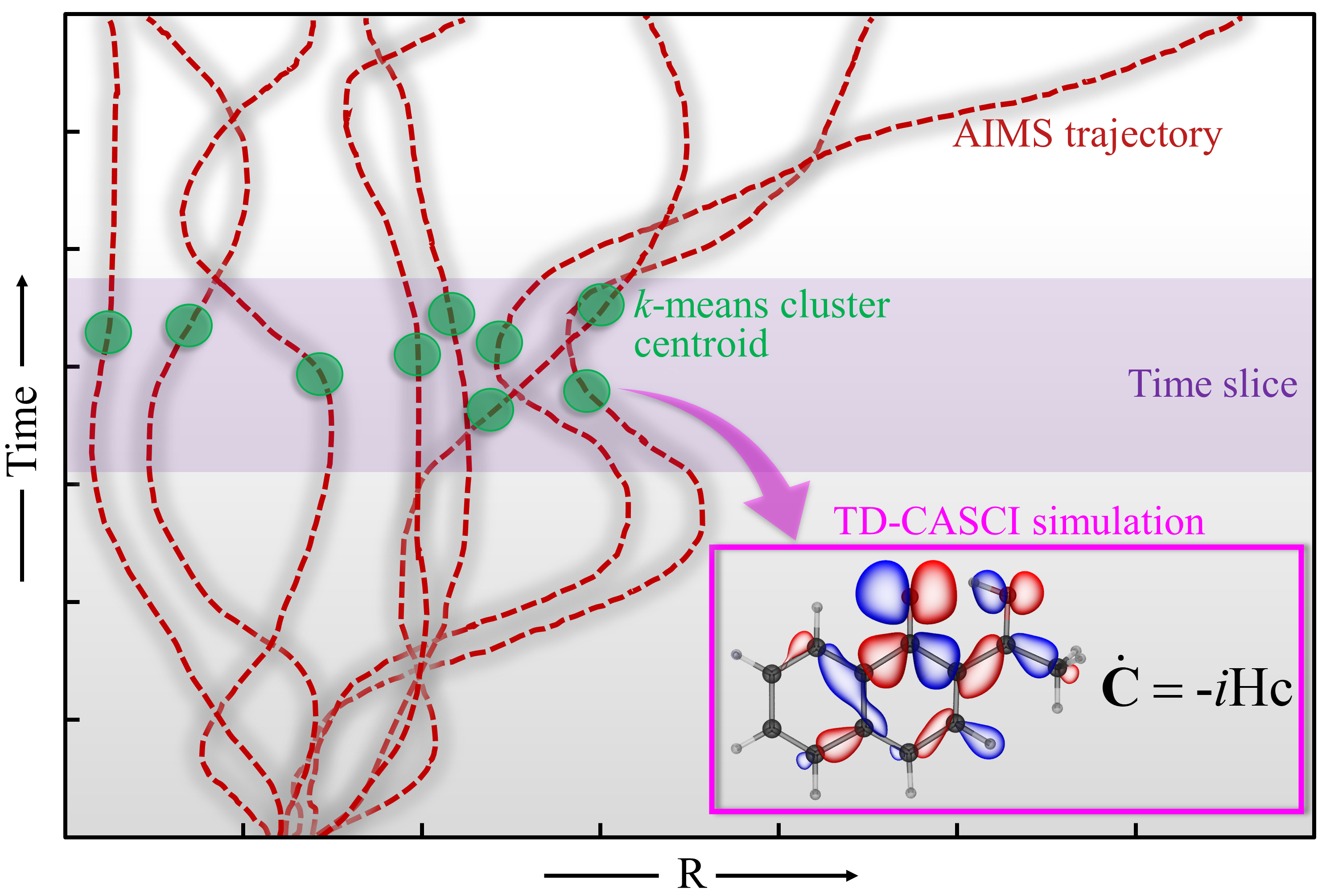}
	\caption{Schematic illustration of TAS simulations through post-processing of NAMD simulations data. The AIMS trajectories (depicted as brown dotted vertical lines) are divided into approximately 12 fs time-slices (represented by light-blue horizontal boxes). From each time-slice, representative geometries (indicated by green circles) are selected as centroids using weighted k-means clustering. For each selected geometry, TAS is simulated by employing 100 fs of TD-CASCI electronic dynamics with the incident field  polarized separately along the x, y, and z directions of the molecular axis.}
	\label{fig:protocol}
\end{figure}

The transient absorption spectrum is simulated in a series of four steps, illustrated in Figure \ref{fig:protocol} :
\begin{enumerate}
\item An approximation to the time-dependent molecular wave function, $\Psi(\mathbf{r},\mathbf{R},t)$, is generated via nonadiabatic \textit{ab initio} molecular dynamics simulations (red dashed lines).  
\item The data is divided into time slices, and each time slice is reduced to a fixed number of representative centroid geometries (green dots) via the weighted \textit{k}-means clustering algorithm.  
\item The ESA and SE spectra corresponding to each centroid geometry are computed via electronic structure calculations at the TD-CASCI level.
\item The spectra of each centroid are summed to generate the full spectrum.
\end{enumerate}  
Each step is described below.

\subsection{Nonadiabatic Dynamics Simulations}

Nonadiabatic molecular dynamics simulations were performed using the ab initio multiple spawning (AIMS) method \cite{Bennun2001,Martinez2002,Martinez2018} as implemented in the PySpawn software package \cite{Fedorov2020}.  Details of the algorithm are presented in the references in the preceding sentence.  Here we only review the basic form of the AIMS molecular wave function as relevant to the task of computing the TAS spectrum.  The AIMS wave function is expanded in a basis of Born-Oppenheimer states,
\begin{equation}
\Psi(\mathbf{r},\mathbf{R},t) = \sum_I \chi_I(\mathbf{R},t) \psi_I(\mathbf{r};\mathbf{R}).
\end{equation}
Here $\psi_I(\mathbf{r};\mathbf{R})$ is the time-independent adiabatic electronic wave function, $I$ indexes adiabatic electronic states, $\mathbf{R}$ and $\mathbf{r}$ are the nuclear and electronic coordinates, respectively, and $t$ is time.  The time-dependent nuclear wave function, $\chi_I(\mathbf{R},t)$, is expanded in a time-dependent set of trajectory basis functions,
\begin{equation}
\chi_I(\mathbf{R},t) = \sum^{N_I(t)}_i c^I_i(t) \chi^I_i(\mathbf{R}; \bar{\mathbf{R}}^I_i(t), \bar{\mathbf{P}}^I_i(t), \boldsymbol{\alpha}^I_i).
\end{equation}
Here $\chi^I_i(\mathbf{R}; \bar{\mathbf{R}}^I_i(t), \bar{\mathbf{P}}^I_i(t), \boldsymbol{\alpha}^I_i)$ is a time-dependent frozen Gaussian \cite{Heller1975} trajectory basis function, whose average position and momentum, $\bar{\mathbf{R}}^I_i(t)$ and $\bar{\mathbf{P}}^I_i(t)$, evolve according to classical equations of motion.  The time-dependent expansion coefficients are propagated via the time-dependent Schrodinger equation.  It is the incoherent weights of the individual basis functions, $|c^I_i(t)|^2$, and their time-dependent average positions, $\bar{\mathbf{R}}^I_i(t)$, that will be used in calculating the TAS spectrum.  In practice, coherences are rare and extremely short-lived in AIMS simulations of systems that relax through conical intersections, thus the effect of including coherences in property calculations is very small.

The electronic wave function, $\psi_I(\mathbf{r};\mathbf{R})$, is computed at the floating occupation molecular orbital (FOMO-) CASCI level of theory.\cite{Slavicek2010,Granucci2001}  An active space of 10 electrons in 10 orbitals is used.  A FOMO temperature of 0.10 a.u. and Gaussian statistics were used to define the occupation numbers.  A 6-31G** basis is used.  We abbreviate this method FOMO(0.10)-CAS(10,10)-CI/6-31G** going forward.  All electronic structure calculations were performed with the TeraChem GPU-accelerated electronic structure software package.\cite{Seritan2021,Ufimtsev2009b,Fales2015,Hohenstein2015}  The accuracy of the chosen active space and basis was established by comparison to accurate complete active space second-order perturbation theory (CASPT2) calculations and equation of motion coupled cluster singles and doubles (EOM-CCSD), as discussed in subsection \ref{sub:validation}.

In this work, the spectrum is averaged over 42 AIMS simulations.  Each simulation is initiated as a single trajectory basis function whose position and momentum, ($\bar{\mathbf{R}}(0)$ and $\bar{\mathbf{P}}(0)$), are sampled from the ground state vibrational Wigner distribution.  The Wigner distribution was defined in the harmonic approximation, using the ground state geometry and frequencies computed at the B3LYP/6-31G**\cite{Lee1988,Becke1993} level of theory.

\subsection{Geometry Clustering}

One could proceed by computing the ESA/SE signals at each computed geometry, but this would be very expensive, because the simulated trajectories contain 184,121 distinct geometries.  To reduce the computational cost, we used a clustering algorithm to select representative geometries.  To this end, the \textit{S}\textsubscript{1} geometric data was binned into 168 sampling windows according to their simulation time.  Each window spans $\Delta t =$ 500 a.u ($\sim$12.1 fs). The conformations belonging to each window were clustered using the weighted \textit{k}-means clustering algorithm to identify \textit{k} $=$ 80 clusters.  The weighted \textit{k}-means clustering algorithm was used as implemented in the SciKit Python package.\cite{SK2011}  Each of the 80 clusters is characterized by a mean geometry, 
\begin{equation}
\bar{\mathbf{R}}_\eta = \frac{1}{w_\eta} \sum_{Ii}^{\text{in cluster }\eta} |c^I_i(t)|^2 \bar{\mathbf{R}}^I_i(t), 
\end{equation}
and a weight, correspond to the total population of the trajectory basis functions assigned to that mean,
\begin{equation}
w_\eta = \sum_{Ii}^{\text{in cluster }\eta} |c^I_i(t)|^2,
\end{equation}
where $\eta$ indexes clusters.  Pairwise root mean squared distance (RMSD) matrices used for clustering were calculated with the MDTraj \cite{McGibbon2015} library.  For each cluster, we define the centroid to be the data point nearest to the mean geometry.  Using this structure to compute the ESA/SE spectra avoids abnormalities resulting from averaging geometries in Cartesian coordinates.  Each of the cluster centroids was translated and rotated using the Kabsch algorithm \cite{Kabsch1976} to minimize the RMSD between the centroid and \textit{S}\textsubscript{0} minimum enol geometry oriented along the \textit{z}-axis of its transition dipole. A total of 13,440 representative time-separated conformations was obtained for the TAS simulations.  

In the present work, only population on S$_1$ was included in the spectrum, because accurate propagation on S$_0$ is unwieldy due to electronic structure convergence issues.  In principle, population on all electronic states can be included by binning population both by time and by electronic state, as was done in our previous work on salicylideneaniline.\cite{Silfies2023} 

\subsection{Simulation of ESA/SE}
The ESA and SE spectra of each centroid geometry were computed by TD-CASCI.  In total, more than 40,000 individual spectra were computed and summed to produce the TAS spectrum of HAN, corresponding to the 13,440 cluster centroids determined above, each probed with three different polarization directions.

In TD-CASCI, the time-dependent electronic wave function is expanded  
\begin{equation}
    \Psi^{CAS}(t) = \sum_{K\in CAS} C_{K}(t) \Phi_{K},
\end{equation}              
where $\{\Phi_{K}\}$ is the set of Slater determinants that define a complete active space (CAS) basis, and $\{C_{K}(t)\}$ is the set of time-dependent, complex CI vector coefficients \cite{Peng2018,Lestrange2018,Liu2019}  Note that in time-dependent configuration interaction, unlike time-dependent self-consistent field methods, the orbitals are not time-dependent, thus the time dependence of the wave function is entirely represented by $\{C_{K}(t)\}$. We leverage the GPU-accelerated direct CI implementations in the TeraChem software package to efficiently integrate the time-dependent Schrodinger equation,\cite{Fales2015,Peng2018}
\begin{equation} 
\label{eq:schrodinger}
    i\frac{\partial \mathbf{C}(t)}{\partial t}=\mathbf{H}(t) \mathbf{C}(t).
\end{equation}
As discussed in previous publications, our implementation recasts Eq. \ref{eq:schrodinger} in symplectic form and uses a second-order symplectic split operator integrator for its propagation\cite{Peng2018}. Our direct CI implementation allows propagation in the full CAS basis, while avoiding the building, storage, and/or diagonalization of the CI Hamiltonian, or any other data structures of its dimension. 

A time-dependent electric field is applied to the system within the dipole approximation, 
\begin{equation}
    \hat{H}(t)=\hat{H}_{0}-\hat{\boldsymbol{\mu}} \cdot \mathbf{d} E(t),
\end{equation} 
where $\hat{\boldsymbol{\mu}}$ is the dipole moment operator, $\hat{H}_{0}$ is the field-free CI Hamiltonian, $E(t)$ is the scalar electric field strength, and $\mathbf{d}$ is the unit vector in the field polarization direction. 

Having propagated the time-dependent electronic wave function, we compute the ESA and SE spectra using a time correlation function formalism.\cite{Tannor2007,Heller2018}  Our simulations are initiated with the electronic wave function in the populated (S$_1$) state, and the electronic dynamics are initiated by a discretized $\delta$-function pulse at time zero.  That is, the pulse is non-zero for only a half time step. The time correlation function is then computed according to
\begin{equation}
    R_x(t)=\mathbf{C}_x(0)^{\dagger} \mathbf{C}_x(t),
\end{equation}
where time zero is the time at the end of the pulse, and $x$ indicates the field polarization direction. The ESA and SE spectra are obtained from $R_x(t)$ according to,
\begin{equation}
R_x(\omega) \propto \omega \mathcal{F} [R_x(t)],
\end{equation}
where $\mathcal{F}[f(t)]$ indicates the Fourier transform of $f(t)$.  We leave out prefactors that are not $\omega$-dependent, because we will only compute the spectrum in relative units.  To facilitate accurate numerical propagation and subsequent interpretation, prior to propagation we shift the zero of energy to the energy of the initial state.  With this shift, signals at positive $\omega$ correspond to ESA, while signals at negative $\omega$ correspond to SE.  The total spectrum, as a function of positive $\omega$, may then be computed by summing the positive SE signal and negative ESA signal according to
\begin{equation}
\Delta S_x(\omega) = R_x(\omega) + R_x(-\omega).
\end{equation}
Note that this neglects ground state bleach signal, which is not present in the current case, because the ground state does not absorb in the probe window.

For each centroid, we perform TD-CASCI calculations with electric fields polarized in the $x$, $y$, and $z$ directions.  Simulations are performed with molecules oriented such that the pump (S$_0 \rightarrow$ S$_1$) transition dipole moment is oriented along the $z$ axis at time zero of the AIMS simulations.  As such, the angularly averaged signals\cite{Hochst2001} corresponding to experiments in which the probe signal is oriented parallel to the pump, perpendicular to the pump, and at a magic angle to the pump (MA) may be computed, respectively, according to
\begin{equation}
\label{eq:para}
\Delta S_\parallel(\omega)=\frac{1}{15} \left[\Delta S_x(\omega)+\Delta S_y(\omega)+3\Delta S_z(\omega)\right],
\end{equation}
\begin{equation}
\label{eq:perp}
\Delta S_\perp(\omega)=\frac{1}{15} \left[2\Delta S_x(\omega)+2\Delta S_y(\omega)+\Delta S_z(\omega)\right],
\end{equation}
and
\begin{equation}
\Delta S_{\text{MA}}(\omega)=\frac{1}{9} \left[\Delta S_x(\omega)+\Delta S_y(\omega)+\Delta S_z(\omega)\right].
\end{equation}

As with time-independent CASCI calculations, orbital selection is an important determinant of the accuracy of TD-CASCI simulations.\cite{Levine2021}  In this work, the FOMO-CASCI method is again used, with the same 10-electron/10-orbital active space and 6-31G** basis as was used in the AIMS simulations.  The FOMO temperature was increased slightly to 0.15 to circumvent some orbital convergence difficulties.  The electronic spectrum of each centroid was derived from a 100 fs TD-CASCI simulation using a time step of 3 as.  Each 100 fs TD-CASCI simulation required $\sim$150 s on a single NVIDIA A100 GPU.  (For comparison, the spectrum simulated using only 45 fs electronic dynamics is given as supporting information Fig. S1.)  

The TD-CASCI electronic dynamics were initiated by a $\delta$ function pulse, approximated as a constant electric field for a 1.5 as (half of a time step) and a field strength of $2.85\times10^4$ a.u ($1.0\times10^{24}$ W/cm\textsuperscript{2}). The choice of pulse intensity is a matter of computational convenience, not a reflection of any experimental reality.  Nonlinear excitation is not possible given the instantaneous nature of our pulse, and therefore a linear spectrum is observed despite the apparently large intensity of the pulse. 
Application for only a single half time step ensures linear response to the field.
Simulations were run for pulses polarized separately along the \textit{x}, \textit{y} and \textit{z} directions. To reduce the effects of spectral leakage, the Hanning windowing function was applied to the raw $R(t)$ prior to Fourier transformation. Taking into account broadening due to windowing, the acquired spectra exhibit a spectral resolution of 0.051 eV ($\Delta\lambda$ = 10.8 nm at $\lambda$ = 400 nm; $\Delta\lambda$ = 33.2 nm at $\lambda$ = 700 nm).  Note that it is not the simulated pulse width that bounds the spectral resolution, but the total propagation time following the pulse. 

\subsection{Summation of Total TAS Signal}

Given the set of spectra corresponding to the centroids of each cluster, $\{\Delta S_\eta\}$, weights of each cluster, $\{w_\eta\}$, and the associated time, $\{t_\eta\}$, we may now sum the total TAS signal according to
\begin{equation}
\Delta S_x(\omega,t) = \sum_\eta w_\eta\ \Delta S_{x,\eta}(\omega) \delta(t-t_\eta).
\end{equation}
Here $x$ may refer either to a probe pulse polarization direction in the theory frame ($x$, $y$, or $z$, where the pump transition is oriented along the $z$ axis) or to the probe pulse orientation relative to the pump (MA, $\parallel$, or $\perp$).  In order to allow direct comparison of the experimental and theoretical data, the theoretical TAS was subjected to processing procedures mirroring the experimental setup. Initially, the theoretical TAS was sampled at regular intervals of 15 nm, matching the experimental point spacing used to record the experimental TA trace. Subsequently, the sampled dataset underwent additional convolution with a Gaussian function characterized by a FWHM of 200 fs in the time domain, matching the experimental time resolution.

In total the simulated spectrum was derived from $\sim$4 ns of TD-CASCI electron dynamics data ($\sim$40,000 simulations, each 100 fs in duration).  The total computational cost, running on NVIDIA A100 GPUs, was $\sim$1,700 GPU-hrs, which is markedly less than the cost of the AIMS simulations from which the spectra were computed.

\section{Experiment}
\label{sec:expt}

The basic principles of cavity-enhanced ultrafast spectroscopy have been described and demonstrated in references \citenum{Reber_Optica2016}, \citenum{Chen_Thesis2018} and \citenum{Allison_JPhysB2017}.
The measurements reported here were conducted using the broadband spectrometer reported by Silfies et al.\cite{Silfies_PCCP2021} with further details in reference \citenum{Silfies_Thesis2023}.
For all measurements, HAN was purchased from Sigma Aldrich inc.\ and loaded into the molecular beam system without further purification.

\begin{figure}[t]
    \includegraphics[width=3.25in]{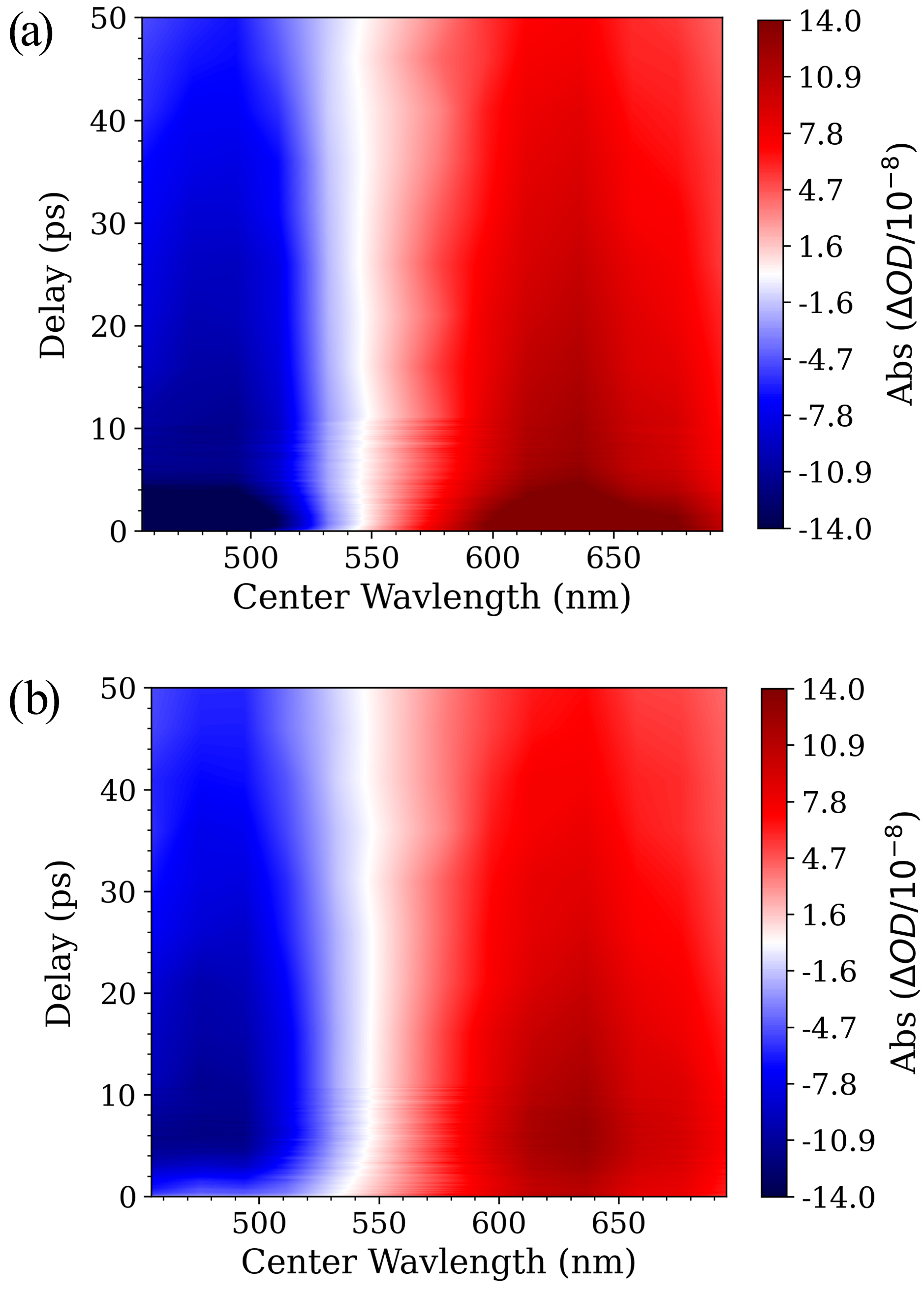}
	\caption{The experimentally measured CE-TAS spectrum with the probe pulse polarized a) parallel to and b) perpendicular to the pump pulse.}
	\label{fig:polar}
\end{figure}

Transient absorption data after excitation at 355 nm were taken for 12 discrete probe wavelengths between 450 and 700 nm, and the broadband spectra displayed here were constructed via interpolation. 
We estimate the instrument response function (cross-correlation between pump and probe) via fitting the rising edges of the signals at each wavelength with an error function, and find a time resolution of approximately 200 fs (FWHM), varying slightly with wavelength.\cite{Silfies_PCCP2021}
For each wavelength, three pump/probe delay scans were recorded with parallel polarization of the pump and probe pulses, and three scans were recorded with perpendicular polarization, and the averages, $\Delta S_{\|}(\lambda)$ and $\Delta S_{\perp}(\lambda)$, were calculated.
Magic angle signals were then constructed with the standard expression $\Delta S_{\text{MA}}(\lambda) = \left[\Delta S_{\|}(\lambda)+2 \Delta S_{\perp}(\lambda)\right] / 3 $.
Polarization-resolved spectra are shown in Figures \ref{fig:polar}a) and \ref{fig:polar}b), and magic angle spectra will be shown below.
The polarization anisotropy observed in this molecule is consistent with the pump and probe transition dipoles being approximately parallel when the subtleties of the CE-TAS signal construction are accounted for as discussed by Silfies et al.\cite{Silfies_PCCP2021}
Global analysis\cite{vanStokkum_GlobalAnalysis2004} of the experimental signal indicates the signal is well fit with a single spectral component with lifetime of 65 ps, similar to the lifetime observed in the previous solution-phase TAS experiment of Lochbrunner, et al.\cite{Lochbrunner2005}

\begin{figure}
	\includegraphics[width = \columnwidth]{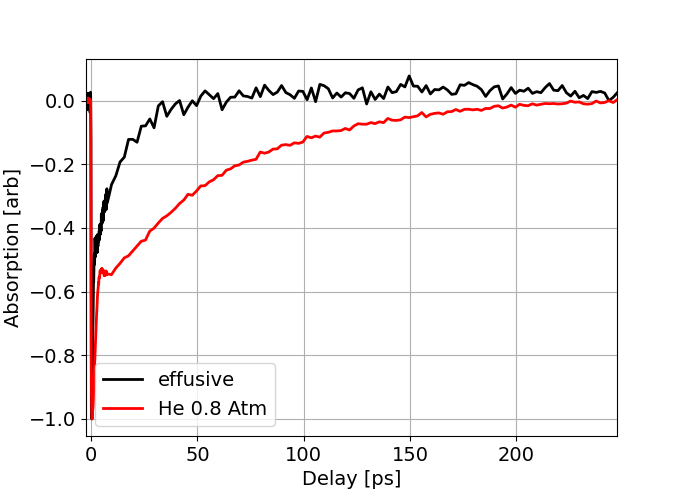}
	\caption{
		Pump/probe stimulated emission signals recorded in HAN at $\lambda = 494$ nm for an effusive molecular beam (black) and the case of 0.8 bar stagnation pressure (red).
		For both signals the pump and probe polarizations are parallel.
		The ``spike'' at early delays is due to the initial rotational coherence produced by the pump pulse.
	}
	\label{fig:hotcold}
\end{figure}

The molecular beam environment of CE-TAS allows us to widely vary the molecular temperature.
Low temperatures in the range of 40-80 K are achieved by flowing He through the slit nozzle with typical stagnation pressures between $\sim$0.1 and $\sim$1 bar.\cite{Silfies_PCCP2021,	Silfies_Thesis2023, Silfies2023}
Using no He flow, we achieve an effusive jet in which the molecular temperature is close to that of the heated nozzle at $\sim$400 K.
In figure \ref{fig:hotcold} we compare the signals from these two conditions.
Immediately apparent is that the excited-state lifetime observed by TAS has a strong temperature dependence, as previously observed in temperature-dependence fluorescence measurements.\cite{Organero2003}
Fitting the longer delay data with a single exponential decay (after the rotational transient which persists for the first 5-10 ps), we obtain time constants of $\tau_{\text{hot}} = 12$ ps and $\tau_{\text{cold}} = 67$ ps.
This strong temperature dependence to the picosecond-scale dynamics of internal conversion in HAN indicates a barrier to the internal conversion on the excited state.
In what follows we use these data to produce an experimental estimate of the barrier height.

To estimate the vibrational temperature of the molecule on the S$_1$ surface after photoexcitation, we consider three factors: (1) The initial temperature and ground-state vibrational energy of the HAN molecules in the molecular beam, (2) the vibrational energy imparted by photoexcitation at 355 nm, and (3) thermalization on the $S_1$ surface after photoexcitation but before internal conversion. 
For (1) we analyze the rotational anisotropy with the formalism developed by Felker,\cite{Felker_JPhysChem1986,Felker_JChemPhys1987} to estimate rotational temperatures of the HAN molecules to be 40 K in the jet-cooled beam and 403 K in the effusive beam, and assume these temperatures to also be the the initial vibrational temperatures in the ground state. 
We use these temperatures to estimate the initial vibrational energies to be 11.5 cm$^{-1}$ and 3583 cm$^{-1}$ in the ground state by calculating the vibrational partition function in the harmonic approximation with ground-state vibrational frequencies from Mawa and Panda.\cite{Mawa2021} 
For (2), from the gas-phase spectroscopy data of Douhal et al.\cite{Douhal1993}, we know the origin of the $S_1$ transition to be at $\lambda \approx 388.6$ nm.
Thus we excite $\sim$2400 cm$^{-1}$ above the origin, leading to 2400 cm$^{-1}$ of vibrational energy in addition to the ground state vibrational energy.
For (3), we estimate the temperature on the excited state by rethermalizing the total vibrational energy (ground-state energy + above-origin excitation energy) among the vibrational modes at the excited keto potential minima provided again by Mawa and Panda,\cite{Mawa2021} with the result $T_{\text{cold}} = 322$ K and $T_{\text{hot}} = 503$ K. 

Finally to estimate the barrier height, $E^{\ddagger}$, from the observed rates and these estimated temperatures, we assume simple Arrhenius behavior
\begin{equation}
	\ln\left(\frac{k_{\text{hot}}}{k_{\text{cold}}}\right) = E^{\ddagger}\left[ \frac{1}{T_{\text{cold}}} - \frac{1}{T_{\text{hot}}} \right] \;.
\end{equation}
Using the measured rates of Figure \ref{fig:hotcold} and our estimated temperatures, we arrive at an estimate for the barrier height of 2.9 kcal/mol. 

\section{Results}
\label{sec:results}

We apply our approach to compute the TAS spectrum of HAN, a prototypical ESIPT system (Fig \ref{fig:enolketo}).  Previously, the excited state dynamics of HAN have been studied experimentally using TA spectroscopy, \mbox{TRPES}, and time-resolved fluorescence.\cite{Lochbrunner2005,Allison2021,Lochbrunner2001,Douhal2004} 
In 2001, Lochbrunner et al. used \mbox{TRPES} to study both ESIPT and the subsequent decay of the excited state \cite{Lochbrunner2001}. 
The fastest decay (30 fs) observed in the experiment was assigned to ESIPT. 
In a follow-up solution-phase TAS experiment in 2005, Lochbrunner et al. \cite{Lochbrunner2005} assigned the rise of the SE signal (104-167 fs, depending on pump wavelength) to a subsequent intramolecular vibrational redistribution (IVR) of the resulting electronically-excited keto isomer.  
In the solution-phase TAS experiment, a $\sim$70 ps decay was assigned to the internal conversion of the excited state, attributed to nonadiabatic decay back down to the ground state.
In our CE-TAS measurements from the jet-cooled molecule, we observe similar kinetics.
Similar lifetimes are observed in the time-resolved fluorescence spectrum, as well, where the decay in cyclohexane at 298K is fit to a biexponential with time-constants 39 and 101 ps.\cite{Douhal2004}
The gas phase absorption and fluorescence spectra of HAN have also previously been reported.\cite{Catalan_JACS1993}
Here we will reconsider the assignments of the TAS spectra by direct simulation of the relevant experimental observable from nonadiabatic molecular dynamics simulation data. 

\subsection{Comparison of Experimental and Theoretical TAS Spectra}

\begin{figure*}[t]
	\includegraphics[width=\textwidth]{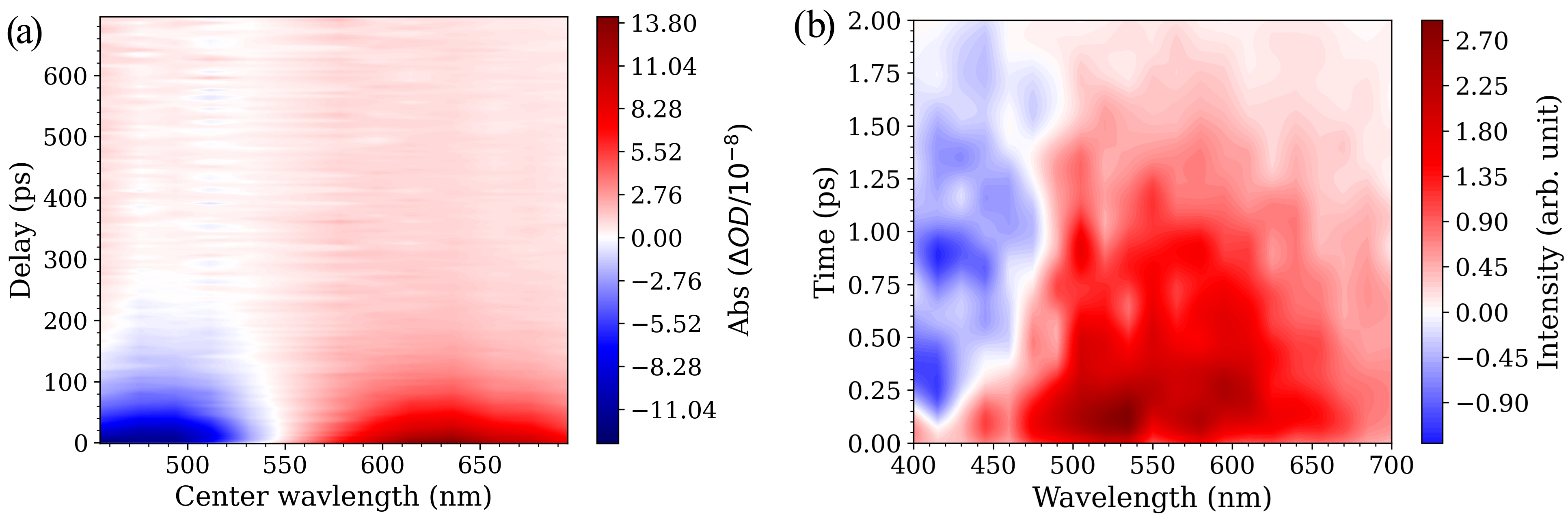}
	\caption{(a) Magic-angle transient absorption spectrum of jet-cooled HAN excited at 355 nm constructed from 12 probe wavelengths. Stimulated emission and excited state absorption are shown using the blue and red colors, respectively. (b) Gas phase TAS simulated at FOMO(0.15)-TD-CAS(10,10)CI/6-31g** level by running the electronic dynamics for 100 fs using the time-resolved geometries of 2 ps AIMS molecular dynamics. The spectrum is constructed using 40,320 individual TD-CASCI simulation by polarizing the field along the \textit{x}-, \textit{y}-, and \textit{z}-directions. Signal intensity, in arbitrary units, for excited state absorption and stimulated emission are indicated with the same colors as used in the experimental plot.}
	\label{fig:combined}
\end{figure*}
\par
Figure \ref{fig:combined} presents a comparison of the experimental magic-angle TAS spectrum of jet-cooled HAN (panel a) with the simulated magic-angle TAS (panel b). ESA is shown in red, while SE is shown in blue. The experimental spectrum for the first 2 ps is provided in Fig. S2 of the Supporting Information. 
The phenomenology of ESA signals overlapping with Stokes-shifted stimulated emission is common for many molecules that undergo ESIPT.\cite{Stock_ChemPhysLett2002,Ziolek_PCCP2004,Zhao_PCCP2012,Chevalier_PCCP2012,Chevalier_JPCA2013}
The experimentally measured and simulated TA have similar spectral features, with an ESA feature observed at longer wavelength, and a SE feature at shorter wavelength. The red edge of the simulated ESA feature is roughly 50 nm shorter than it is in the experiment (at 550 nm).  

In both spectra, the ESA and SE features decay with the same time constant.  However, the simulated lifetime (1.7 ps) is a factor of $\sim$40 shorter than the experimental one (67 ps).  On first glance this discrepancy appears very concerning, but we will demonstrate below that this large error in lifetime can be attributed to a modest error in the PES.  Less error prone than computed lifetimes are computed spectra, which are not exponentially sensitive to errors in computed energies.  Thus, the logic of our analysis going forward will be to assign the spectrum not by comparing experimental and theoretical time constants, but instead by determining which contributions to the simulated spectrum arise from which molecular motions.  Further discussion of the discrepancy between lifetimes once we have assigned the decay of the lifetime to a physical process.

Before continuing, we briefly mention two details the reader should keep in mind while comparing the spectra.  First, the noisy appearance of the simulated spectrum in Fig. \ref{fig:combined}b is an artifact of the finite number of cluster centroids employed (80 per time slice) and the relatively high-energy resolution afforded by 100 fs probe simulations.  Second, as in the experiment, the decomposition of the computed spectrum into parallel and perpendicular components confirms that both the ESA and SE signals are polarized roughly parallel to the initial excitation (Figure S2).

\subsection{Assignment by Decomposition of the Simulated Spectrum}

To assign the spectral features to molecular motions, in this subsection we decompose the spectrum into components corresponding to different dynamical processes.  First, consider the fact that the signal arises as the sum of two contributions: a positive signal from ESA, and negative signal from SE.  Cancellation between ESA and SE complicates interpretation, but theory allows us to view them separately.

\begin{figure*}[t]
	\includegraphics[width=\textwidth]{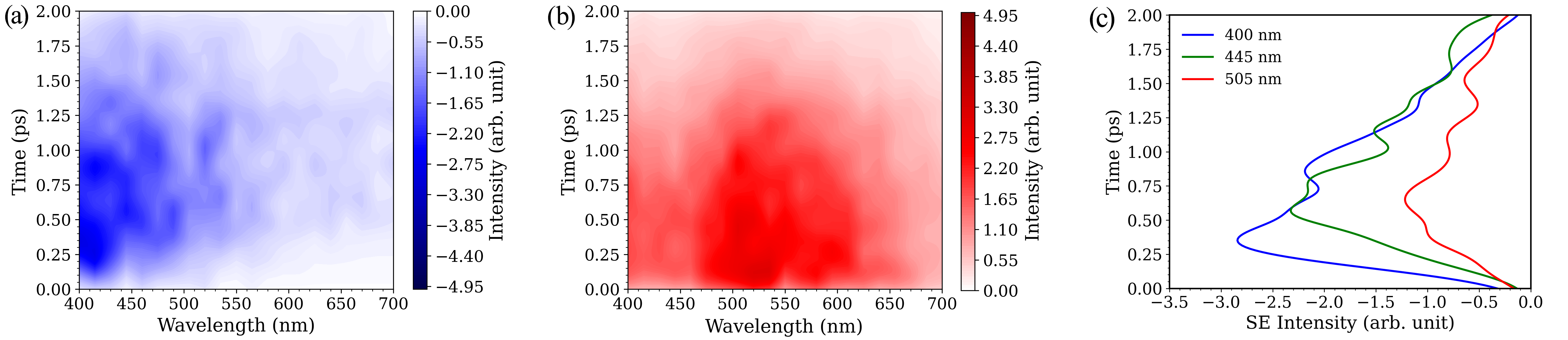}
	\caption{Magic-angle simulated plots of (a) stimulated emission and (b) Excited state absorption of HAN at FOMO(0.15)-TD-CAS(10,10)CI/6-31g** level by running the electronic dynamics for 100 fs using the time-resolved geometries of 2 ps AIMS molecular dynamics.  Panel (c) shows the magic-angle simulated plots of stimulated emission at three specific probe wavelengths as a function of time.}
	\label{fig:ESA_SE_combined}
\end{figure*}

Figure \ref{fig:ESA_SE_combined} shows the separate contributions to the simulated TA spectrum from SE ($R_{MA}(-\omega)$; panel a) and ESA ($R_{MA}(\omega)$; panel b).  Note that they strongly overlap, with a significant portion of the SE signal obscured by more intense ESA at longer wave length.  Note also that the profiles of the two signals behave differently with time.  The ESA signal is large at time zero, and decays over the 2-ps window of the simulations without shifting significantly in frequency.  In contrast, the SE is nearly zero at time zero, within the detection window, and rises in early time.  The rise time varies with wavelength, with longer wavelengths appearing at later times.  This can be seen more clearly in Figure \ref{fig:ESA_SE_combined}c, which presents time slices of the simulated SE signal at three different probe wavelengths.  The maximum SE is observed at 0.3, 0.6 and 0.7 ps for 400, 445, and 505 nm, respectively.  This behavior corresponds to a reduction in the S$_0$-S$_1$ energy gap as the molecular relaxes on the excited state, i.e. the Stokes shift.  This feature is not definitely observed in the present experimental spectrum, given limited time resolution.  However, a rise in the SE between 500 and 560 nm with a slightly faster timescale ($\sim$0.15 ps) was observed in the solution-phase TAS experiments by Lochbrunner, et al.\cite{Lochbrunner2005}, and assigned to relaxation following faster (30 fs) ESIPT.

\begin{figure*}[t]
	\includegraphics[width=\textwidth]{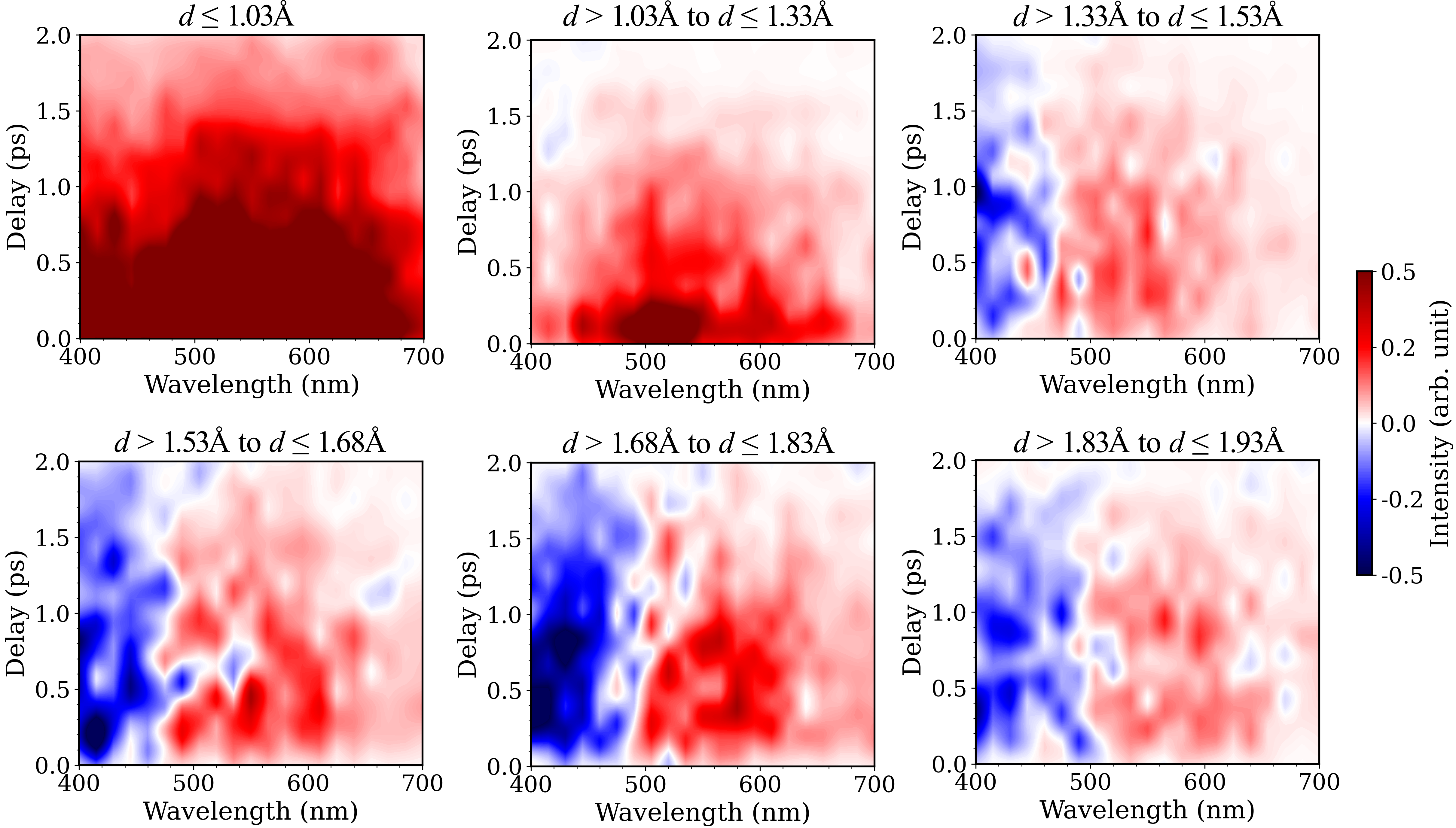}
	\caption{Dynamic development of TAS as a function of ESIPT path. The magic angle TAS is plotted for different O\textsubscript{D}-H bond lengths, where the bond length (\textit{d}; defined in Fig. \ref{fig:enolketo}) increases from left-right and from top to bottom. Note that the O\textsubscript{D}-H bond length in the enol component at FC geometry is 0.99 Å while it is 1.47 Å in the \textit{S}\textsubscript{1} keto minimum geometry.}
	\label{fig:esipt}
\end{figure*}

To revisit this assignment, we decompose our spectrum to separately visualize the contribution of specific degrees of freedom in the evolution of TAS as a dynamic picture. Figure \ref{fig:esipt} shows the TAS spectrum decomposed into contributions from structures that fall at different points on the proton transfer coordinate (defined as the the distance between the donor oxygen atom and the transferring proton, $d$ in Fig. \ref{fig:enolketo}). Each spectrum contains the contribution to the TA spectrum from the set of geometries where $d$ falls within a specific range, regardless of what time those geometries were explored. As such, we may assign spectral contributions to specific chemical species.  In Figure \ref{fig:esipt} we include only relatively planar geometries (acetyl twist angle $\phi \leq$ 30 deg), so that we may separately consider out-of-plane twisting below. Here we see that ESA is observed for all values of $d$, but that the SE feature grows in only when the proton transfers, with the largest contributions arising for $d>1.68$ Å. To provide a complementary view of this information, a video file showing the TAS spectrum "develop," with each component of Fig. \ref{fig:esipt} added one at a time is provided as supporting information.  Note that we see signal from all points along the reaction coordinate at all points in time because our simulated spectrum is derived from a swarm of AIMS simulations in which the proton transfers at different times.  This approximately reflects the quantum mechanical uncertainty in the position of the proton at any particular time.  

Based on this result, it is tempting to assign the rise in the SE feature of the TAS signal to ESIPT, in contrast to the previous assignment to IVR of the keto form, post-ESIPT.\cite{Lochbrunner2005}  But if we are to assign the rise in SE to ESIPT, we must also revisit the previous assignment of the 30 fs decay in the \mbox{TRPES} spectrum to ultrafast ESIPT.\cite{Lochbrunner2001}  Thus we consider the possibility that this 30 fs decay may correspond to the initial relaxation of the enol isomer, preceding ESIPT.  To this end, we performed multi-state complete active space perturbation theory (MS-CASPT2) calculations of neutral S$_1$ and the cation ground state (D$_1$) at the Franck-Condon and the enol S$_1$ minimum energy geometries.  These calculations were performed in OpenMolcas\cite{OpenMolcas} using a ten-electron/ten-orbital active space and the 6-31G** basis.  At this level, relaxation of the enol pre-ESIPT results in a 0.6 eV increase in the D$_1$-S$_1$ gap, which is enough to account for the rapid decay of the \mbox{TRPES} spectrum.  Thus, we conclude that the 30 fs process observed in the \mbox{TRPES} corresponds to relaxation of the enol, pre-ESIPT, and the longer ($\sim$150 fs) rise time in the TAS corresponds to ESIPT itself.

\par
\begin{figure*}[t]
	\includegraphics[width=\textwidth]{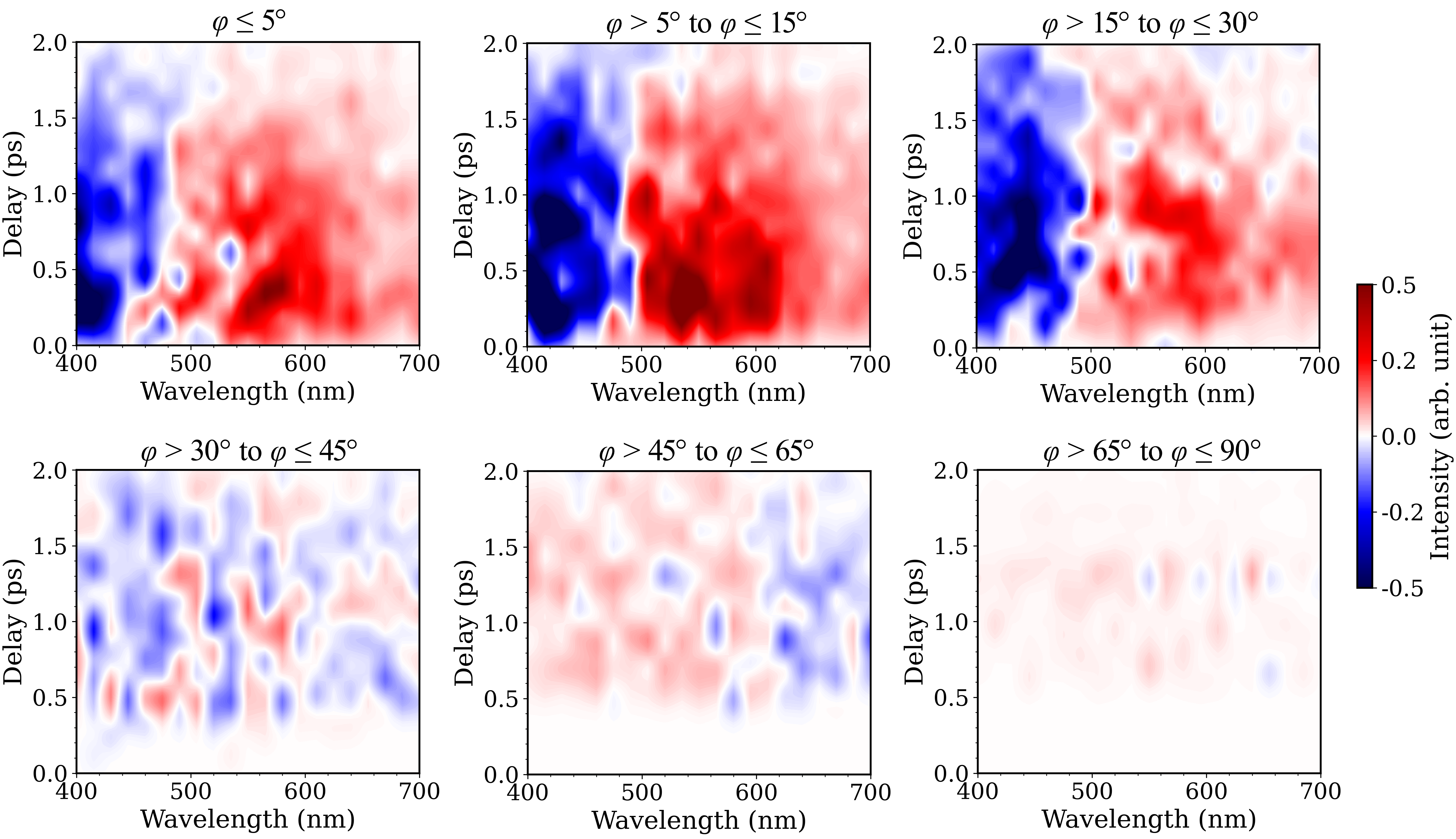}
	\caption{Dynamic development of TAS as a function of acetyl rotation path. The magic angle TAS is plotted for different values of the dihedral angle defined in Fig. 1. The value of the dihedral angle increases from left-right and from top to bottom.}
	\label{fig:dihedral}
\end{figure*}

Following ESIPT, our simulations indicate that the keto isomer undergoes acetyl rotation, in agreement with previous work by Douhal.\cite{Organero2000,Douhal2004}  The TAS spectrum, decomposed into components associated with different acetyl twist angles ($\phi$ in Figure \ref{fig:enolketo}), is shown in Fig. \ref{fig:dihedral}.  At small twist angles ($\phi \leq 30^{\circ}$) the spectrum is relatively insensitive to $\phi$.  And for $\phi > 30^{\circ}$, there is very little contribution to the signal, because the excited state population is rapidly quenched via the CI.  Upon relaxation to S$_0$, both ESA and SE disappear.  Thus, consistent with past assignments, we attribute the slow (~70 ps) decay of the TAS signal to relaxation to the ground state.

The factor of $\sim$40 difference between the experimental and theoretical lifetimes, which looks extremely concerning on its face, can be attributed to a relatively small error in the PES.  In order to twist to reach the CI geometry, the molecule must traverse a transition state.  At the FOMO level of theory used in our molecular dynamics simulations, the barrier height is found to be 0.5 kcal/mol (0.02 eV).  This is considerably smaller than the 2.9 kcal/mol barrier determined from the temperature-dependent CE-TAS data above and the 3.4 kcal/mol barrier determined by Douhal et al. from the temperature dependence of the time-resolved fluorescence of HAN in cyclohexane.\cite{Douhal2004}
Thus we conclude that FOMO underestimates the proton transfer barrier by 2-3 kcal/mol.
 Errors of a few kcal/mol are inevitable when using the low levels of electronic structure theory that are compatible with molecular dynamics simulations, such as FOMO-CASCI.  Yet at a low temperature they translate to very large errors in lifetime.  Our relatively simple choice of zero-Kelvin Wigner initial conditions likely also contributes to the error in the predicted lifetime.

\subsection{Validation of PES} \label{sub:validation}

\begin{figure*}[t]
    \includegraphics[width=\textwidth]{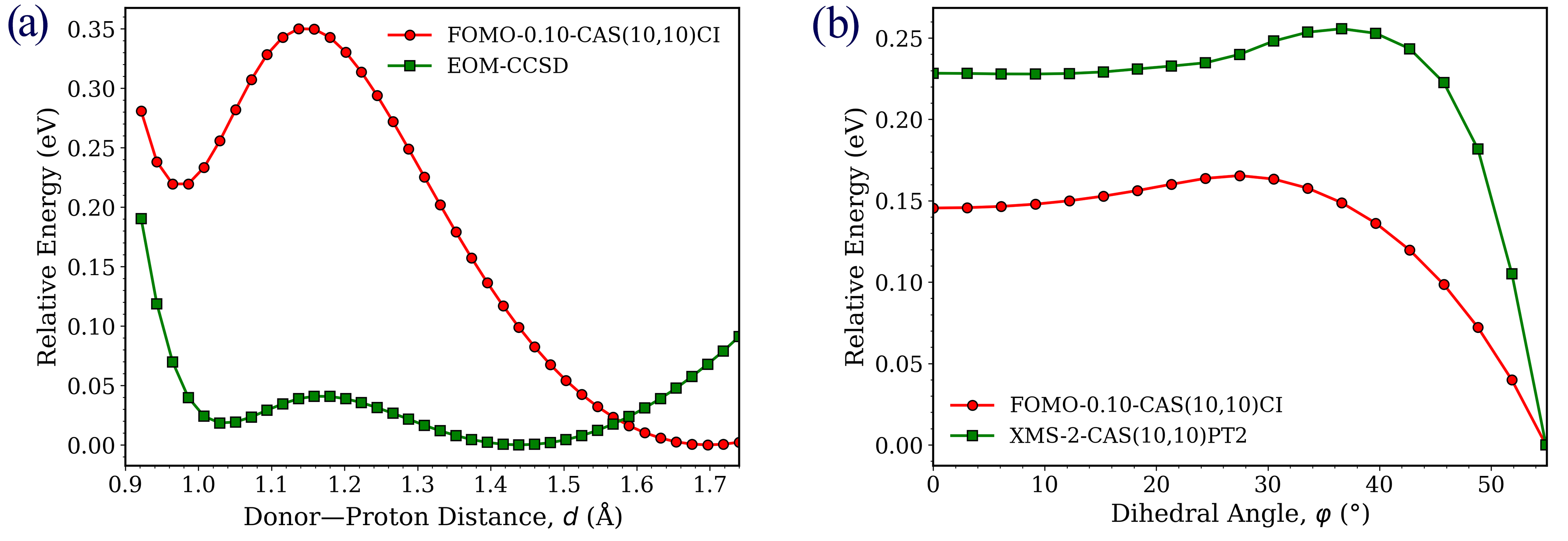}
	\caption{Relaxed S$_1$ PESs following the a) proton transfer and b) acetyl twisting coordinates, computed at the FOMO-CAS(10/10)-CI/6-31G** level of theory.  For comparison, the proton transfer PES is also computed at the EOM-CCSD/6-31G** level, using geometries optimized at the CAM-B3LYP/6-31G** level.  Similarly, the acetyl twisting PES is computed at the XMS-CAS(10/10)-PT2/6-31G** level using geometries optimized at the FOMO-CASCI level.}
	\label{fig:validation}
\end{figure*}

To analyze the accuracy of the FOMO PES used in the molecular dynamics simulations, we computed two relaxed PES scans following important reaction coordinates and compare to higher levels of theory.  Both FOMO scans are computed at geometries determined by constrained optimization at the FOMO-CAS(10,10)-CI/6-31G** level of theory.  The first scan (Fig. \ref{fig:validation}a) follows the proton transfer pathways, comparing the FOMO surface to EOM-CCSD energies computed using time-dependent density functional theory (TDDFT; CAM-B3LYP\cite{camB3LYP}/6-31G**).  The EOM-CCSD and TDDFT calculations were performed using the Psi4 software package.\cite{Psi4}  Both scans show a small barrier to proton transfer, though the FOMO barrier (0.13 eV) is considerably larger than that at the EOM-CCSD level (0.03 eV).  The existence of a barrier to ESIPT is consistent with our assignment of the rise of the SE signal to ESIPT.  In addition, the assessment that FOMO overestimates the proton transfer barrier is consistent with the fact that our simulated timescale for this rise (0.3-0.7 ps) is notably longer than the rise observed in the TAS experiments of Lochbrunner, et al. ($\sim$0.15 ps).\cite{Lochbrunner2005}  The overestimation of proton transfer barriers by CASCI has reported for other systems, as well.\cite{Pijeau2017A,Pijeau2017B}

The second scan (Fig. \ref{fig:validation}b) follows the twisting of the acetyl group as the molecule approaches the conical intersection to the ground state.  The FOMO results are compared to extended multi-state complete active space second order perturbation theory\cite{XMSCASPT2} (XMS-CASPT2) calculations performed using the OpenMolcas software package.\cite{OpenMolcas}  The same 10/10 active space and 6-31G** basis was used as for the FOMO calculations, though a CASSCF reference was used.  The XMS-CASPT2 calculations were performed along the FOMO-optimized reaction path geometries.  Both methods predict a small barrier to acetyl torsion, though the barrier is slightly larger at the XMS-CASPT2 level than at the FOMO level.  This is in qualitative agreement with the suggestion that FOMO underestimates the barrier, as we have previously argued based on the temperature-dependence of the CE-TAS data.

\section{Conclusions}
\label{sec:conclusions}

In this work we have computed the gas-phase transient absorption spectrum of HAN using an efficient and robust approach based on AIMS simulations of ultrafast nonadiabatic dynamics and GPU-accelerated TD-CASCI simulations of the probe.  The primary features of the spectrum closely match those observed experimentally, though the ultimate decay of the simulated signal is a factor of $\sim$40 faster than that of the experiment.  This apparently large discrepancy can be attributed to a relatively modest (2.4 kcal/mol) error in the barrier to acetyl twisting on S$_1$.  In this case, the simulation of the spectroscopic observables (ESA and SE) enable conclusive assignment of this spectral feature to relaxation through a twisted CI, despite the large error in lifetime.

In addition, decomposition of the spectrum into slices associated with different positions along the proton transfer coordinate enables us to assign the rise in the SE signal in the first $\sim$150 fs after excitation to ESIPT.  Static CASPT2 calculations of the ionization potential support the notion that faster (30 fs) timescales observed in a previously reported \mbox{TRPES} spectrum\cite{Lochbrunner2001} can be assigned to relaxation on the S$_1$ PES prior to ESIPT.

This work underlines the utility of direct calculation of ultrafast spectroscopic observables and simultaneous analysis of multiple experimental datasets (e.g. TAS and \mbox{TRPES}) collected under similar gas-phase conditions. In practice, application of both the molecular dynamics and electron dynamics tools used here to nanoscale ($\sim$1-2 nm) systems would be practical. The size of the active space typically determines computational cost, and a small active space is often sufficient even for nanoscale systems.\cite{Levine2021}  However, several limitations exist, providing avenues for further advance.  Since TD-CASCI does not allow excitations to the virtual orbitals outside the CAS space, our current approach is restricted to dynamics within the valence excited electronic states. In future work, we plan to include single excitations to the virtual orbital space to extend our method to high-energy processes. In addition, extension of robust TDCI-based approaches to the simulation of ultrafast x-ray absorption experiments, multi-dimensional spectroscopic measurements, and ultrafast dynamics in solution\cite{Liu2019} will enable a deeper connection between ultrafast experiment and theory.

\section{Supplementary Material}

The supplementary materials for this paper include 1) a document that includes a) the simulated magic angle spectrum computed from shorter (45 fs) TD-CASCI simulations, b) simulated spectra with parallel and perpendicular probe polarization, and c) optimized molecular structures; 2) an hdf5 binary data file containing both the experimental and simulated spectral data; 3) python scripts for plotting the spectra from the hdf5 file.

\begin{acknowledgments}
This work was supported by the National Science Foundation, under grant number CHE-2102319.  AM, ASD, and BGL acknowledge startup funding from Stony Brook University and the Institute for Advanced Computational Science.  MCS and TKA acknowledge support from the Air Force Office of Scientific Research (AFOSR) under grant FA9550-20-1-0259.  This work used Expanse GPU at the San Diego Supercomputer Center (SDSC) through allocation CHE140101 from the Advanced Cyberinfrastructure Coordination Ecosystem: Services and Support (ACCESS) program, which is supported by National Science Foundation grants \#2138259, \#2138286, \#2138307, \#2137603, and \#2138296.
\end{acknowledgments}

\section*{Data Availability Statement}

The data that support the findings of this study are available from the corresponding author upon reasonable request.

%\nocite{*}
\bibliography{aipsamp}% Produces the bibliography via BibTeX.

\end{document}